\documentstyle[aps]{revtex}
\begin{document}
\title{Some Features of the Hadronic $B_c^{(*)}$-meson\\
Production at Large $p_T$}
\author{A.V.~Berezhnoy, A.K.~Likhoded, O.P.~Yushchenko\\
{\it Institute for High Energy Physics, Protvino 142284, Russia\\
E-mail: LIKHODED@MX.IHEP.SU}}
\maketitle

\begin{abstract}
Calculations of the hadronic $B^{(*)}_c$-mesons production
performed in the framework of the perturbative QCD taking into account
$O(\alpha_s^4)$ Feynmann diagrams are presented.
A comparison of the exact calculations with those based on the
fragmentation model of $\bar b\rightarrow B^{(*)}_c+X$
 shows the large discrepancy
between  them. The exact calculations of the $B^{(*)}_c$-mesons production
cross-sections as the function of $p_T$ at the energy of the FNAL Tevatron
($\sqrt{s}=1.8$ TeV) are given. The predicted ratio of the vector to
the pseudoscalar state cross-sections is about $R\sim 3$ instead of
$R\sim 1.4$
for the  fragmentation model.
\end{abstract}

\section{INTRODUCTION}

The production of hadrons with two heavy constituent quarks of different
flavours was the subject of the theoretical studies over two the last
years. The calculations of the $B_c^{(*)}$-meson
production cross-section in $e^+e^-$
annihilation at the $Z^0$ pole [1-3] show a principal possibility
of the ground state observation  of the
$\bar b c$ system as well as its higher
excitations at LEP-I energies. It is much more possible to observe
$B_c^{(*)}$-meson
in experiments at the hadronic collider where we can expect [4-6]
a large number of events at the available luminosity of the FNAL collider.

Another aspect of interest to the processes of the heavy quark system
production consists in the fact that
it offers a possibility to understand the mechanism of the hadronization
in the framework of the perturbative QCD with the minimal assumptions
concerning wave functions of the bound states.

Let us enumerate the main results of the studies of the $\bar b c$ system
production mechanisms in different reactions:

\begin{description}
\item[i)] $B_c^{(*)}$-mesons production in $e^+e^-$ annihilation at high energy
$(M^2_{B_c}/s\gg 1)$ can be described
in the framework of the fragmentation model with the simple factorized
expression
\begin{equation}
\frac{d\sigma_{B_c^{(*)}}}{dz}=
\sigma_{b\bar b}\cdot D_{\bar b\rightarrow B_c^{(*)}}(z),
\end{equation}
where $z=2E_{B_c}/\sqrt{s}$, and $D_{\bar b\rightarrow B_c^{(*)}}(z)$ is the
 fragmentation function of $\bar b\rightarrow B_c$ [1-3].

The ratio of the vector $B^*_c$-meson production cross-section to that
of the pseudoscalar $B_c$ is
\[
R=\frac{\sigma_{B^*_c}}{\sigma_{B_c}}=1.4
\]
instead of the expected value $R\simeq 3$ that one can obtain counting
the quark states.

The production of the $P$-level states in the fragmentation model is
about by an order of magnitude suppressed with respect to that for
 $S$-states [7].

The contribution of the $c$-quark fragmentation into the final cross-section
is about two orders of magnitude lower comparing with that of the $\bar
b$-quark.
\item[ii)] The production of $B_c(B_c^*)$-mesons in the
$\gamma\gamma$ collisions is described by 20 Feynmann diagrams [8-10]
that can be split into three gauge invariant groups (6+6+8 diagrams,
correspondingly). The first two groups describe the process of the
$b\bar b$ and $c\bar c$ pairs production with the subsequent
fragmentation into $B_c^{(*)}$-meson. The contribution of the
$c$-quark fragmentation
in $\gamma\gamma$ collisions
is enhanced due to the factor $(Q_c/Q_b)^4$ and can't be neglected contrary
to the case of $e^+e^-$ annihilation.

The contribution of the $\bar b$-quark fragmentation can be sufficiently well
described by equation (1) and the accuracy of this description increases
with the growth of $p_T$ value.
The contribution of the $c$-quark fragmentation is in a quite drastic
contradiction with factorized expression (1).
The main contribution into $B_c(B_c^*)$-meson production at any $p_T$
comes from the remaining 8 diagrams of the recombination type.
This results in another value of $R>3$ different from that in $e^+e^-$
annihilation.
\item[iii)] The hadronic production of  $B_c^{(*)}$-meson
requires the calculation
of the full set of 36 Feynmann diagrams of the order of $O(\alpha_s^4)$
including fragmentation type diagrams.
The cross section of the $B_c(B_c^*)$ can be then obtained by the convolution
of the gluon-gluon cross-section with the gluon distribution functions in the
initial hadrons. This leads to the dominance of the region of small
$\sqrt{\hat s}$ of the sub-process
$gg\rightarrow B_c^{(*)}+\bar c$ in the cross-section of $B_c$ production.
Just in the same way as in $\gamma\gamma$ collisions the recombination type
diagrams dominate in $gg$ production of $B_c(B_c^*)$ [6] and, consequently,
the $R$ value is close to 3 in the hadronic production as well.
\end{description}

It is obvious that the experimental cuts on $p_T$ values rises up the role of
large $\hat s$ in the reaction $gg\rightarrow B_c^{(*)}+X$.
In such a case we should expect the increase of the fragmentation mechanism
contribution and, consequently, the simplification of all calculations.
The calculations at large $p_T$ values have been performed in [5,11]
where the cross-sections of $B_c(B_c^*)$-meson production for the FNAL Tevatron
are presented. The authors of the cited papers  used as the fragmentation
function for $\bar b\rightarrow B^{(*)}_c$ the value obtained in $e^+e^-$
annihilation
for the $S$ and $P$ levels. Then, convoluting this fragmentation function
 with the cross-section of the
$b\bar b$ pair production they calculated the distribution of
$B_c(B_c^*)$-mesons as the function of $p_T$. The leading logarithmic
correction on the
final state gluon radiation was also taken into account.

In this paper, considering the full set of Feynmann diagrams of the order of
$O(\alpha_s^4)$  we will show that the fragmentation approach fails,
because not all the diagrams, which can't be neglected at large $p_T$, were
taken into account and there was some overestimation of the real phase
space.

We will give a more detailed analysis of the
$B_c(B_c^*)$ production at large $p_T$ for the energy of the FNAL Tevatron
($\sqrt{s}=1.8 \mbox{ TeV})$.

\section{HADRONIC PRODUCTION OF $B_c^{(*)}$ WITH LARGE $p_T$}
It seems obvious at  the naive level that the dominating mechanism
of the $B_c^{(*)}$-mesons production at large $p_T$ should be connected with
the fragmentation of heavy $\bar b$-quarks. The fragmentation approach is
based on the assumption that the factorization of the production on the
parton $(\bar b)$ production with large energy and its subsequent fragmentation
into different  $B_c$ states is valid. The differential cross section
$d\sigma/dp_T$, for example, for $\bar p p$ collision will have the form
\begin{eqnarray}
\frac{d\sigma}{dp_T}&(&\bar p p\rightarrow H(p_T)x)=
\sum_{i,j}\int dx_1dx_2dz f_{i/p}
(x_1,\mu)f_{j/\bar p}(x_2,\mu)\times\nonumber \\
\times \frac{d\hat\sigma}{dp_T}
&(&ij\rightarrow \bar b(p_T/z)+x)\times D_{\bar b\rightarrow H}(z,\mu),
\end{eqnarray}
where $D(z,\mu)$ are fragmentation functions $\bar b\rightarrow H;\,
H=B_c, B^*_c...,\,
d\hat\sigma/dp_T$ is $b \bar b$ pair production cross-section and
$f_{i/A}(x,\mu)$ is parton $i$ density function in the hadron $A$.
This approach was used in [5,11], where the differential
distributions $d\sigma/dp_T$ for  $B^{(*)}_c$-meson and its $P$-wave excitation
were calculated.
Nevertheless, it remains unclear what is the region where we can use
expressions (2) and what can be the role of other subprocesses.
Let us clarify these points in the framework of the same approximation --
the Born-level diagrams of the order of $O(\alpha_s^4)$ [6].
The total number of diagrams in the subprocess
$gg\rightarrow B_c^{(*)}+X$ is 36 and only minor part of them is connected with
the $\bar b$-quark fragmentation.

We have already showed [6] that the main contribution into $B^{(*)}_c$-meson
production is connected with the diagrams of the recombination type, where
both initial gluons dissociate into a pair of heavy quarks and two of these
quarks
recombine  later into $B^{(*)}_c$-meson.
We observed the dominance of such diagrams up to the energies of
$gg$-collisions
of the order of 1 TeV. Moreover, at the energies
$\sqrt{\hat s}\sim 30$ GeV, that give the main contribution into hadronic
production of $B^{(*)}_c$-meson it is rather meaningless to consider the
fragmentation mechanism at all, because the condition  $M^2/\hat s \ll 1$
is not valid and the pre-asymptotic terms in the definition of $D(z)$ are
large. So, we can use the fragmentation approach at large values of
$\sqrt{\hat s}$ only, but at large energies the fragmentation
 contribution is small as compared with that of the whole set of diagrams.

In Fig.~1 we present the results of the
exact calculation for the cross-section of
the
subprocess $gg\rightarrow B_c(B^*_c)+X$ as a function of c.m.s. energy
in comparison with that calculated in the fragmentation approach --
the cross-section of the $b\bar b$ pair production multiplied by the
probability of the fragmentation $\bar b\rightarrow B_c$ and
$\bar b\rightarrow B^*_c\,\, ,\; 3.\cdot 10^{-4}$ and
$4.15\cdot 10^{-4}$, correspondingly.
One can see, that in the regions where the condition
$\hat s/M^2_{B_c}\gg 1$ is valid and expressions (2) can be used
the contribution of the fragmentation is not dominant.
Contrary, at small energies it gives overestimated values due to
the fact that in the fragmentation calculation one uses two-particle
phase space ($2\rightarrow 2$) instead of true three-particle space
($2\rightarrow 3$) in the partonic subprocess.

The same conclusion can be drawn from Fig.~2,
 where we present the differential
cross-sections $d\sigma/dp_T$ for $B_c$- and $B^*_c$-mesons
in comparison with the fragmentation mechanism at the energy
$\sqrt{\hat s}=100$ GeV in gluon-gluon collision.
In the case of $B_c$-meson production we observe the saturation of the
fragmentation at the values of $p_T >30$ GeV, while in the case of
$B^*_c$-meson this saturation is postponed to the very edge of the
phase space ($p_T>40$ GeV).

To stress the problems with the fragmentation picture let us consider the
differential cross-section  $d\sigma/dp_T$ for  $B_c$- and
$B^*_c$-mesons at the energy of gluon-gluon collision   $\sqrt{\hat s}=20$
 GeV (Fig.~3).
One can see that incorrect calculation of the phase space leads to
the fact that
the cross-section calculated by  fragmentation expressions (1)
exceeds that obtained from the full set of diagrams.

The conclusion one can draw from the above considerations is quite simple:
at the currently reachable $p_T$ the fragmentation approach does not work
and all the diagrams contributing into  $B^{(*)}_c$ mesons should be
considered.

It is important to underline another point. The full and the fragmentation
calculations give different predictions for the ratio of the cross-sections
of $B_c$ and $B^*_c$ productions. The full calculation gives the value of
$R\simeq 3$, while the fragmentation model predicts another value
 $R\simeq 1.4$.

In Fig.~4 we present the differential cross-section
$d\sigma/dp_T$ for $B_c$ and
$B^*_c$-mesons calculated from $gg$ cross-section
convoluting it with structure functions of the initial protons for the
energy of the FNAL Tevatron $(\sqrt{s}=1.8$ TeV). We take
the structure functions from [12] fixing virtuality at the value of
 $Q\sim 10$ GeV, because we consider only those
Born diagrams contributions, which
 approximately correspond  to those virtualities $(\alpha_s\simeq 0.2)$.
The running virtuality and the strong coupling constant can be used when
one considers next-order radiative corrections to this process.

Two curves in the same figure correspond to the contributions calculated
from expression (2) using fragmentation functions
$D_{\bar b \rightarrow B_c}(z)$ and $D_{\bar b \rightarrow B^*_c}(z)$
obtained at the
same order of perturbative expansion.
 One can see that those curves
do not coincide with the exact calculations in the whole interval of $p_T$.
At small values of $p_T$ one can observe excess of  fragmentation
predictions, while at large values of $p_T$ they are lower.

Taking into account the experimental cuts on the pseudo-rapidity
($|y|<1$) one can calculate more realistic $p_T$ distribution (Fig.~5).
Such a cut reduces the value of cross-section by a factor of 3.
It is interesting to note that in the $p_T$ interval considered the value
$R=\sigma_{B_c^*}/\sigma_{B_c}$ is about 3. Taking into account all
$S$-wave excitations the cross-section of the $B_c$-meson production
$1S (B_c+B^*_c)+2S(B_c+B^*_c)$ with $p_T>5$ GeV is about 3.3~nb
that gives for Run Ib of  Tevatron with the integrated luminosity
 $100\div 150\mbox{ pb}^{-1}$ about
$3.3\div 5.0 \cdot 10^5 \,\, B_c$-mesons.

\section{DISCUSSION}

There are two different approaches to the calculation of the hadronic
production
of $B_c^{(*)}$-mesons at present. The first one is based on the
calculation of all the diagrams of the order of $O(\alpha_s^4)$ in the
perturbative QCD. There are three publications with such  calculations, which
give different predictions for  cross-sections [4, 6, 13].
In our previous work [6], where we considered the hadronic $B_c^{(*)}$
production,
we had omitted the color factor  $1/\sqrt{3}$ in the wave function of
 $B_c^{(*)}$-meson
that increased our prediction by a factor of 3.
Correcting this point and using the same constants as in [4] we have obtained
a good agreement at the level of the $gg$ cross-sections.

The second approach is based on the usage of the fragmentation model
[5,11], and, as we have demonstrated above, it gives incorrect description of
the
cross-section underestimating it at large energies of $gg$ collisions and
 overestimating at small values of energies.
So, it looks quite strange that the authors of [11] when using fragmentation
approach and convoluting the resulting expressions with the structure functions
have obtained the  cross-sections, which  are analogous to ours (see Fig.~5).

Finally, we would like to note:

\begin{description}
\item[1)] In the case of the states with equal mass heavy quarks
(say, $\psi$ production)
the fragmentation approach $c\to \psi$ does not also describe the
exact result obtained by the calculating of all the diagrams of the order of
$O(\alpha_s^4)$.
\item[2)] When one quark becomes light ($u, d, s$ instead of $c$-quark)
in the region of large $p_T$ the recombination diagrams become dominant.
\end{description}

All these points we will consider in our next publications.

\section*{REFERENCES}
\begin{enumerate}
\item
          L.Clavelli, {\em Phys. Rev.}, {\bf D26}(1982), 1610;

          C.-R.Ji and R.Amiri, {\em Phys. Rev.}, {\bf D35}(1987), 3318;

          C.-H.Chang and Y.-Q.Chen, {\em Phys. Lett.}, {\bf B284}(1992), 127.
\item
 E.Braaten, K.Cheung and T.C.Yuan, {\em Phys. Rev},
 {\bf D48}(1993), 4230.
\item
 V.V.Kiselev, A.K.Likhoded and M.V.Shevlyagin, {\em Z.Phys},
{\bf C63}(1994), 77.
\item
C.-H.Chang and Y.-Q.Chen, {\em Phys. Rev.}, {\bf D48}(1993), 4086.
\item
K.Cheung, and T.C.Yuan, {\em Phys. Lett.}, {\bf B325}(1994), 481.
\item
A.V.Berezhnoy, A.K.Likhoded and M.V.Shevlyagin, {\em Preprint}
IHEP-94-48 (1994), hep-ph/9408284.
\item
T.C.Yuan, {\em Phys. Rev. Lett.}, {\bf 71}(1993), 3413.
\item
A.V.Berezhnoy, A.K.Likhoded and M.V.Shevlyagin, {\em Phys. Lett.},
{\bf B342}(1995), 351.
\item
 K.Ko\l odziej, A.Leike and R.R\"uckl, {\em Preprint} MPI-Ph.T/94-84,
LMU-23-94 (1994).
\item
F.Sartogo and M.Masetti, {\em private communication}.
\item
K.Cheung and  T.C.Yuan, {\em Preprint} UCD-95-4 and CPP-94- (1994).
\item
J.Botts et al., ``CTEQ Parton Distributions and Flavour Dependence
of the Sea Quarks'', {\em Preprint} ISU-NP-92-17, MSUHEP-92-27 (1992).
\item
 S.R.Slabospitsky, {\em Preprint} IHEP-94-53 (1994).
\end{enumerate}

\section*{FIGURE CAPTIONS}
\begin{itemize}
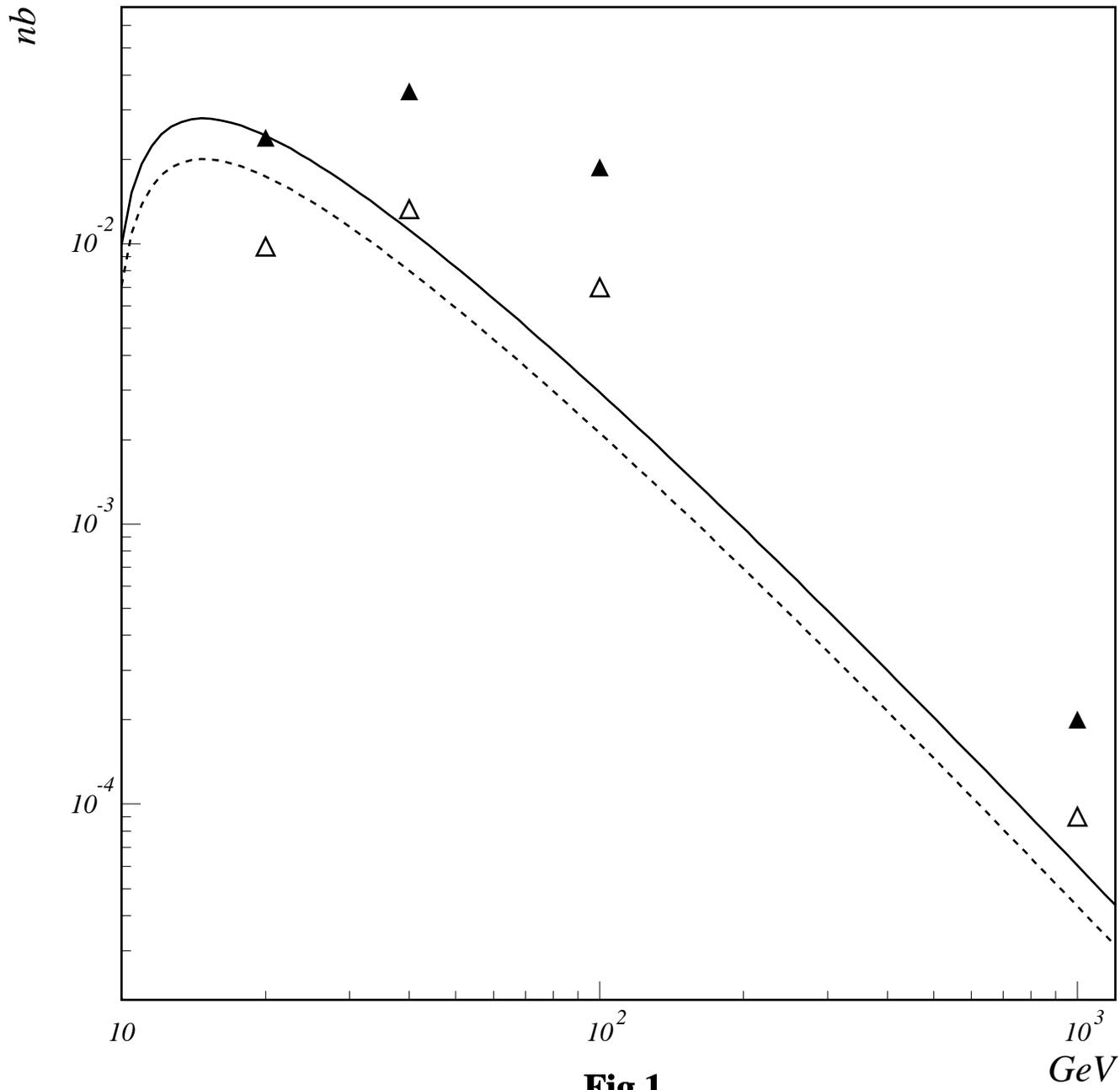
\item[Fig.1.]
Gluon cross-section in nb for the $B_c$ (white triangle)
and $B_c^*$ (black triangle) production. The fragmentation model predictions
for
$B_c$ (dashed line) and $B_c^*$ (solid line) are given for comparison.

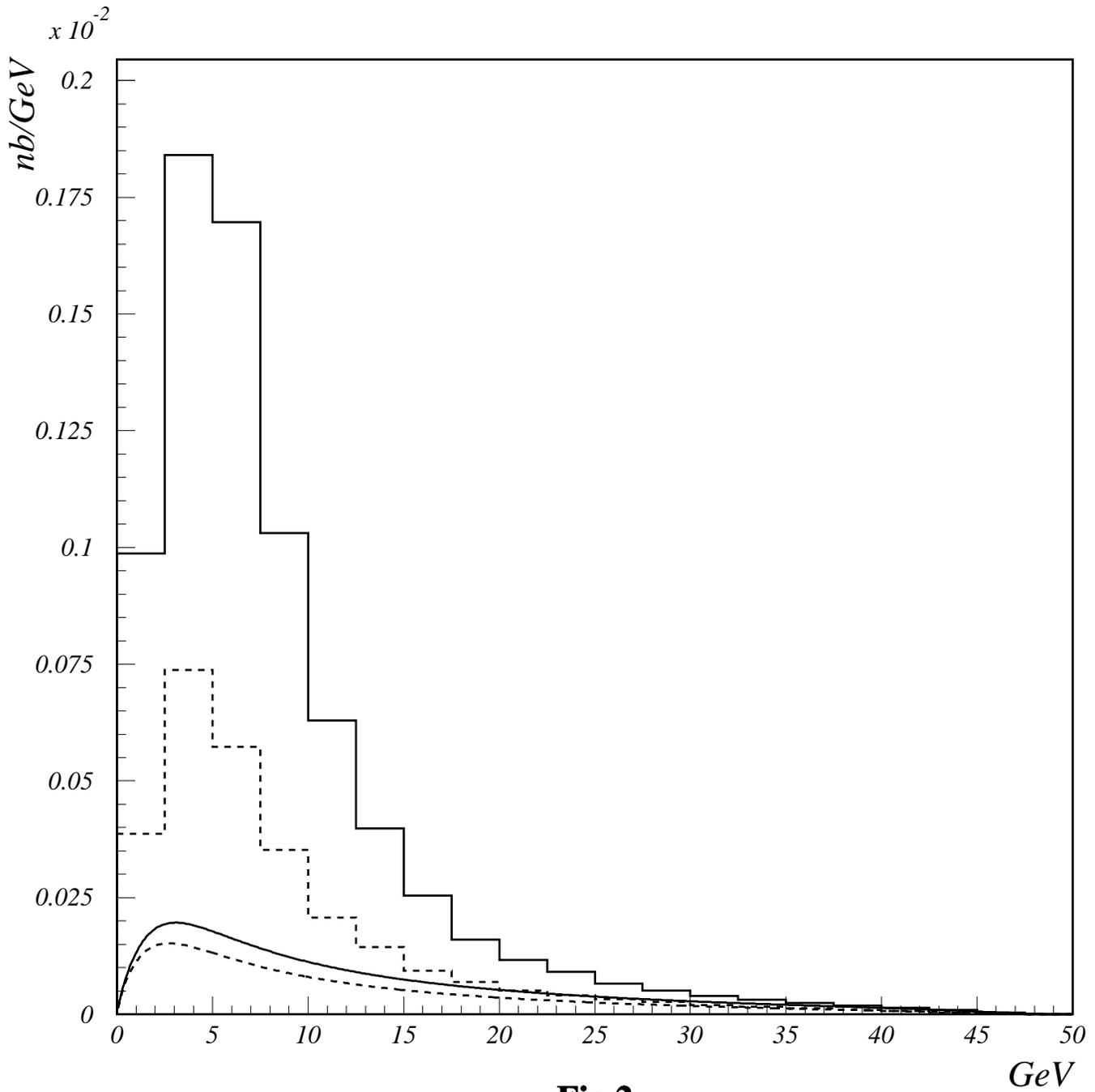
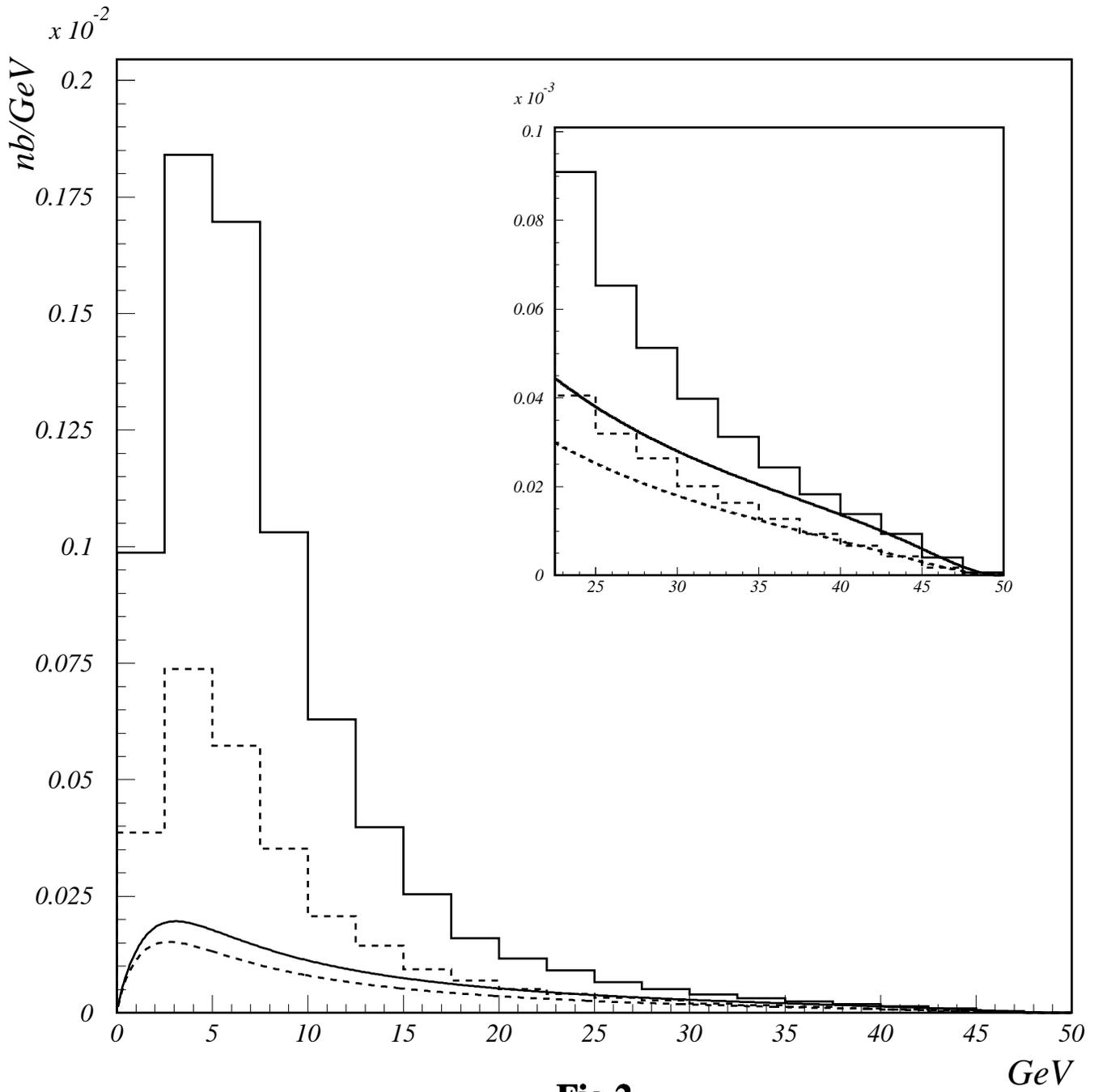
\item[Fig.2.]
$d\sigma/dp_T$ distribution in $nb/GeV$ of $B_c(B_c^*)$ production
(histograms)
in $gg$ collisions in
comparison with the fragmentation approach (curves) at the energy of 100 GeV.

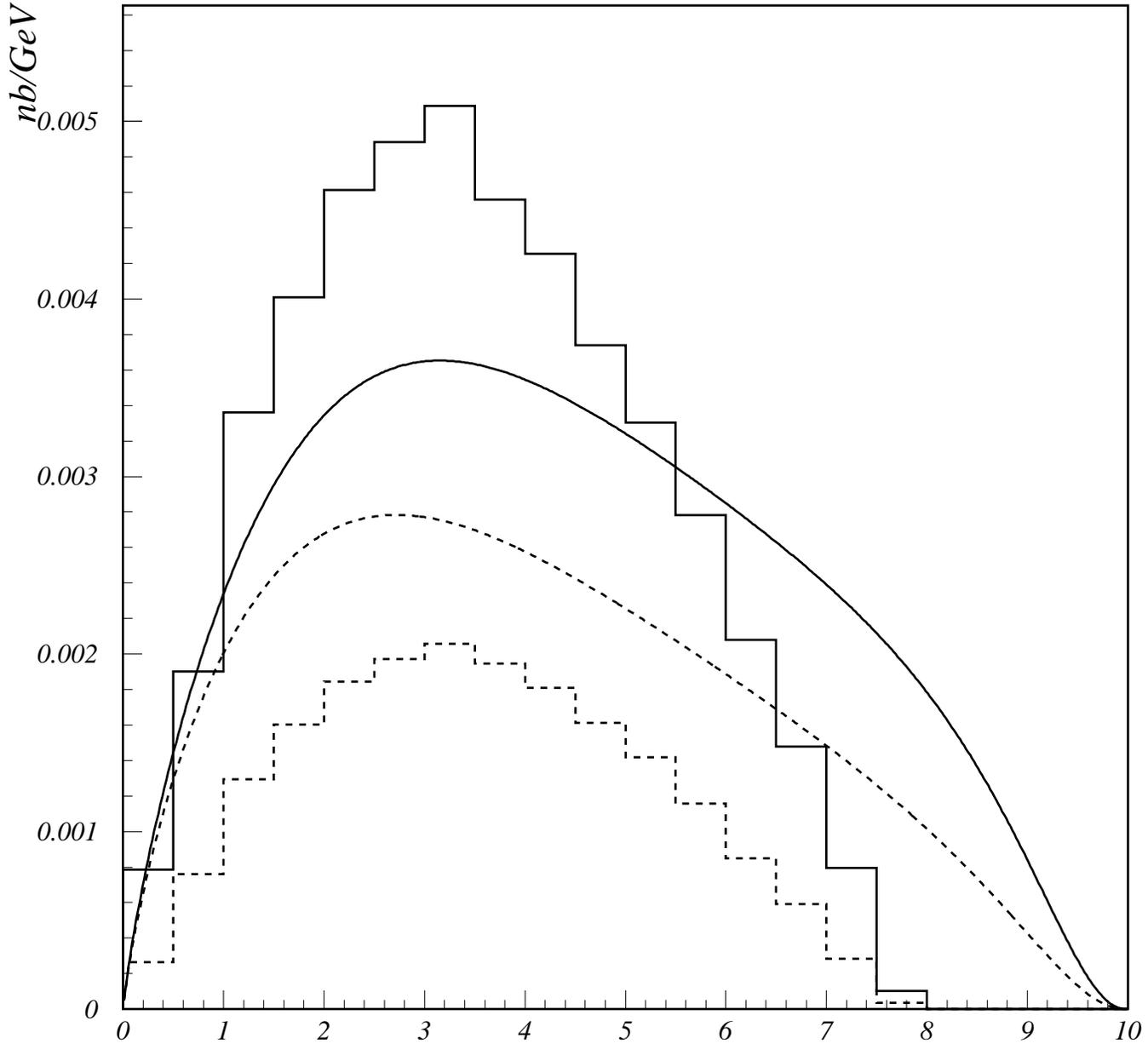
\item[Fig.3.] The same as in Fig.2, but for the energy of 20 GeV.

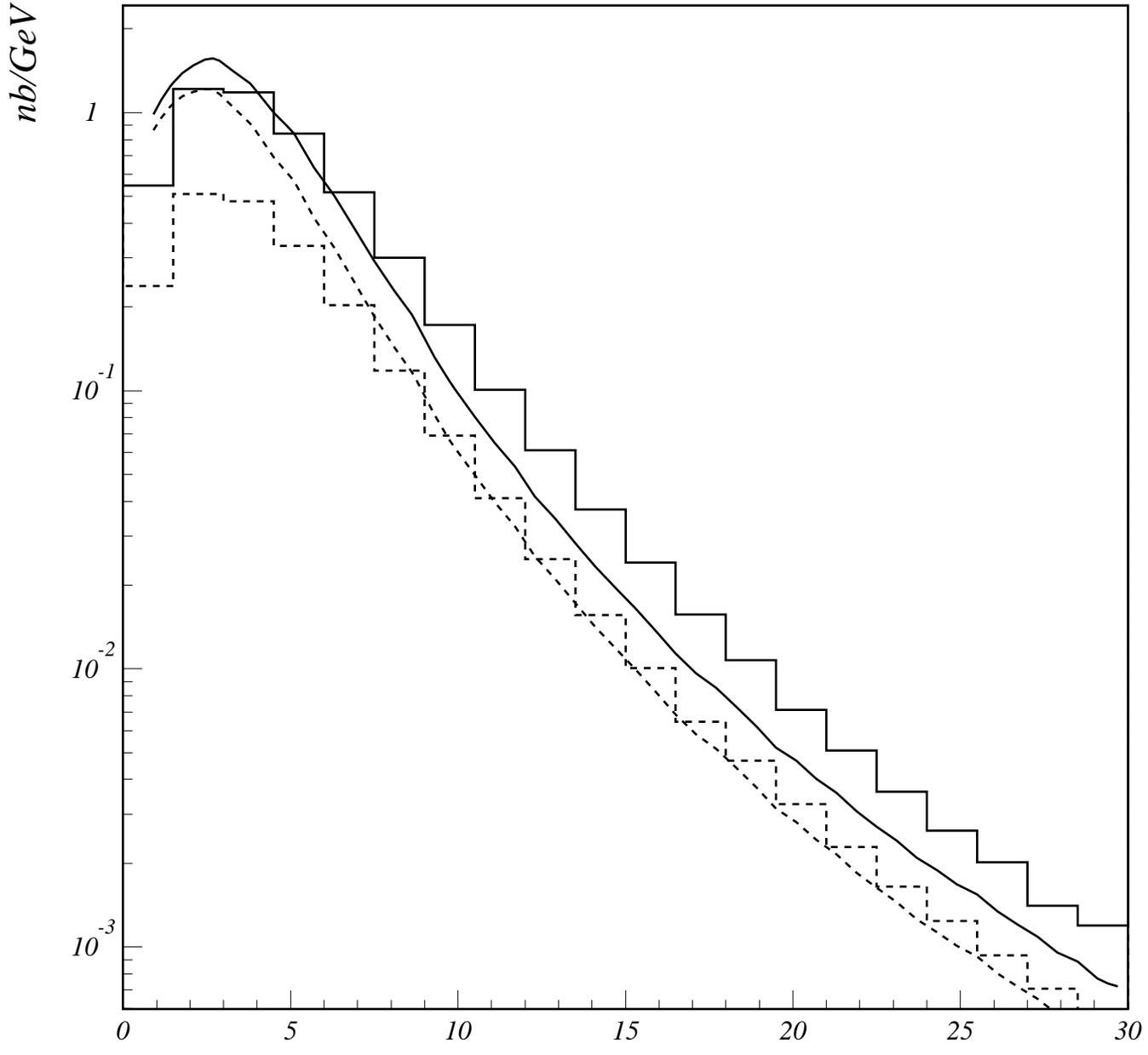
\item[Fig.4.] Differential cross-section
$d\sigma/dp_T$ for $B_c(B_c^*)$ in $p\bar p$-collisions at $\sqrt{s} = 1.8$ TeV
in comparison with the fragmentation approach.

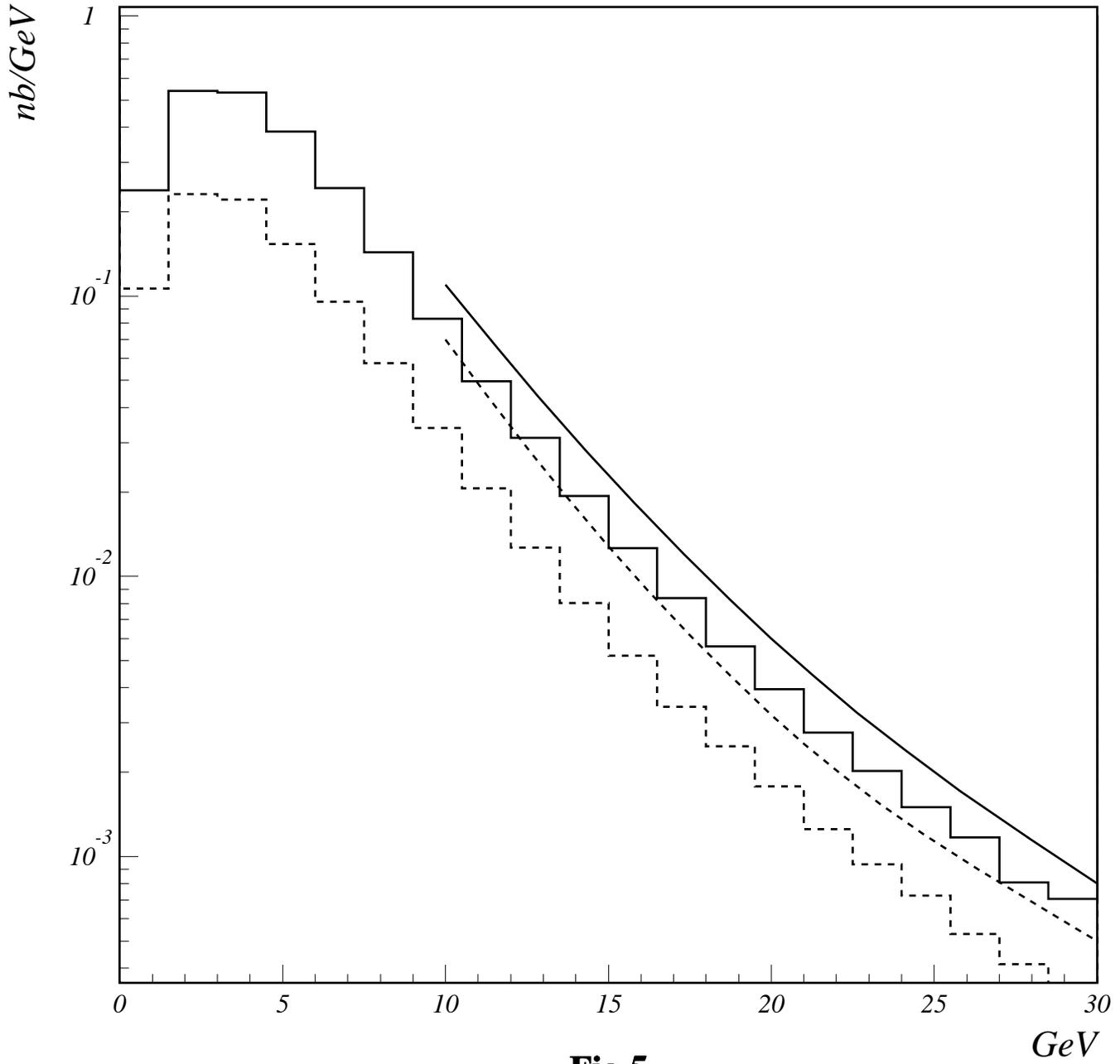
\item[Fig.5.] The same
 as in Fig.4, but with cut $|y(B_c^{(*)})|<1$. The curves correspond to the
predictions from [11].
\end{itemize}
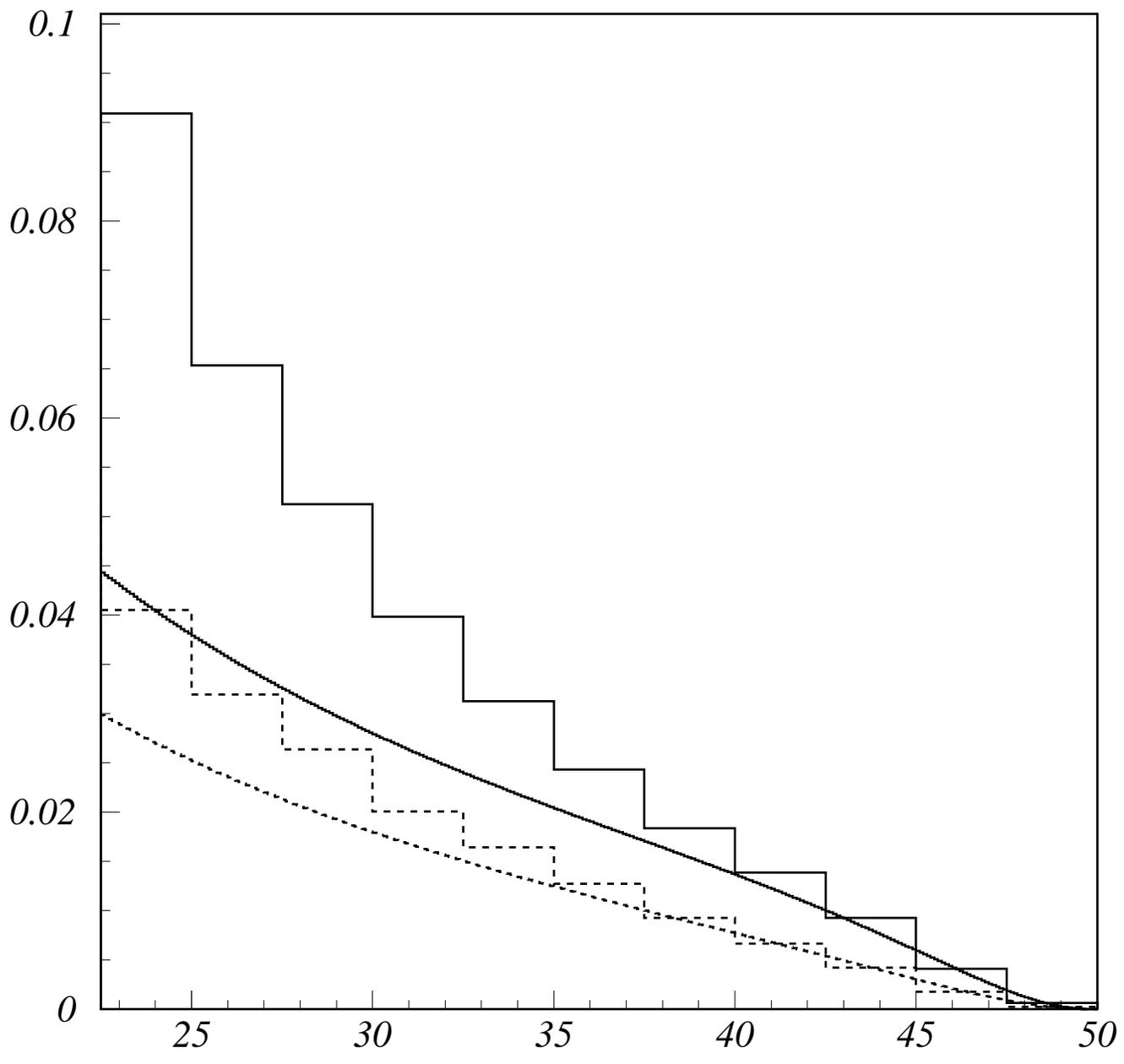
\end{document}